# Impact of core – cladding boundaryshape on the waveguide properties of hollow core microstructured fibers

A. D. Pryamikov[a)], G.K. AlagashevandA. S. Biriukov

*Fiber Optics Research Center of Russian Academy of Sciences, 119333, Moscow, Russia*

In this paper we consider an interaction between the air core modes (ACMs) of hollow core waveguide microstructuresandcore – cladding boundary walls in various shapes. The analysis is based on well - established models such as the ARROW (anti-resonant reflecting optical waveguide) model and on the models proposed for the first time. In particular, we consider the dynamics of light localization in the polygonalcore cladding boundary wall asdependant on the type of its discrete rotational symmetry. Based on ourfindingsweanalyzethe mechanisms of light localization in the core – cladding boundary walls of negative curvature hollow core microstructured fibers (NC HCMFs).

**I. INTRODUCTION**

By now, a large variety of waveguide microstructures from holey fibers [1] to hollow core fibers with complicatedmicrostructured claddings [2 - 4] has been developed. Corresponding mechanisms of light localization from a modified total internal reflection to a photonic band gap mechanism [1] have been proposed. The general approach to explaining the light localization mechanism in HCMFs is called the ARROW mechanism [5]. The ARROW mechanism implies that periodicity in the photonic crystal cladding does not always give rise to the Bragg constructive interference necessary for the light localization in the core. If the wavelength of the localized radiation is short enough in comparison with the unit cell of the photonic crystal cladding $\lambda < \Lambda$ each high index layer can be treated as an individual scatterer. Each of these layers can be considered as a planar Fabry – Perot resonator with its own resonance system while the impact of the cylindrical surface can also be taken into account [6]. In this case, the transmission band edges are determined by the transverse resonances of these individual cladding layers but not the Bragg resonances of the cladding. Inside the bands, light reflects effectively from each of the layers in an antiresonant regime and interferes in the air core while forming low leaky modes.

Recent studies have shown that two following factors play the key role in the process of light localization in HCMFs with a complicated cladding structure: the structure of a photonic crystal cladding [3, 4] (its antiresonant properties) and resonant properties of the hollow core which are determined by its size and the shape of the core – cladding boundary. For HCMFs with a photonic crystal cladding [3] and a circular core – cladding boundary, the light localization in the hollow core occurs as a result of the spectral overlapping of the resonances of the hollow core and theantiresonant spectral range of the cladding [1] (photonic band gaps). For HCMFs with a Kagome lattice photonic crystal cladding, the light localization mechanism is

_________________________

[a)] pryamikov@mail.ru.

more complicated and not based on the photonic band gaps. At present, a generally accepted explanation of the waveguide mechanism in HCMFs with a Kagome lattice photonic crystal cladding is based on the inhibited coupling model [7, 8]. According to this model, the light is localized in the hollow core due to the low density of optical states in the Kagome lattice photonic crystal cladding and to a small value of the overlap integral between the air core modes and the cladding states.

In [9] a HCMFwith a Kagome lattice photonic crystal cladding and a core – cladding boundaryin the form of a hypocycloid curve was demonstrated. In [10] the same authors demonstrated the possibility of a drastic decrease in the total loss in a HCMF with a Kagome lattice photonic crystal cladding by means of hypocycloid core – cladding boundary with avaryingdegree of the negative curvature.

It is clear from the above considerations that the structure of the microstructured cladding (in particular, its antiresonant properties) and the resonant properties of the hollow core determined by its size and shape play a key role in the process of light localization in HCMFs.

For the first time, the pronounced effect of the core – cladding boundary shape on optical properties of HCMFs was demonstrated and investigated in [11, 12]. In these works the waveguide structures didn't have a complicated photonic crystal cladding. In particular, in [11] the authors considered themicrostructured cladding consisting of eight silica glass capillaries located symmetrically on the inside surface of silica glass mounting tube. In [11] the HCMFswith a periodic change of theinner normal direction tothe core – cladding boundarywere called 'negative curvature' HCMFs.In [12] the silica glass microstructured cladding had a core – cladding boundary in the form of joint "parachutes" (an "ice – cream cone" in the authors' terminology). Despite the simplification of the HCMF cladding structure the transmission bands for the air core modes were observed experimentally up to wavelengths of 7 – 8 μm [13], where the silica glass is opaque. In [14] the impact of thenegative curvature of the core – cladding boundary on optical properties of HCMFs was analyzed numerically. Also, the calculations showed that even in the case of the simplest NC HCMF structure consisting of eight silica glass capillaries [11] 99.993% of radiation directed by the fiber can propagate in air [15]. It can then be concluded that the material properties of the cladding don't play such an important role in the formation of the ACMs as geometrical parameters of the core – cladding boundary.

In this work, we have considered hollow core waveguide microstructures with a complicated shape of the core - cladding boundary to analyze the impact of geometrical parameters of the core – cladding boundary on optical properties ofHCMFs. Some of them have a polygonal shape of the core – cladding boundary with an azimuthal unit cell $\delta\varphi = 2\pi / N$, where $N$ is a number of the polygon sides. The sides of other polygonal waveguide microstructures have a 'negative curvature'. Based on the analysis of the behavior of ACM fields in the core - cladding boundary wall it is possible to provide insights into the mechanism of light localization in NC HCMFs. As it will be shown, the difference between the light localization mechanisms in HCMFs with a polygonal shape of the core - cladding boundary (including NC HCMFs) and in HCMFs with a circular shape of the core – cladding boundary lies in different waveguide antiresonant regimes for the ACM fields in the core – cladding boundary walls. In particular, the different antiresonant regimes for the ACM fields in the polygonal microstructures walls are also observed at different values of $\delta\varphi = 2\pi/N$. Moreover, the transverse electric field distributions of the ACMs of different orders vary in the polygonal microstructure walls within the transmission band at a given wavelength. As it will be shown it occurs due to more complicated behavior of the ACM fields in the polygonal microstructure walls with a determined type of discrete rotational symmetry compared to the one in the wall with continuous rotational symmetry. Correspondingly, it leads to a modification of the standard ARROW mechanism of light localization [5, 6] and to a decrease in the waveguide loss. It also will be shown that the light localization mechanism in the negative curvature waveguide microstructures is similar to that in waveguide microstructures with a polygonal shape of the core – cladding boundary wall.

The rest of the paper is divided into 3 parts: Section 2, where we consider the difference between the processes of light localization in waveguide microstructures with a polygonal shape of the core – cladding boundary wall and the circular dielectric tubes; Section 3, where we consider the mechanism of the ACMs formation in NC HCMFs based on the results of Section 2; and Section 4 where we give the conclusions.

## II. Light localization in waveguides with a polygonal shape of the core – cladding boundary

We start by comparing the mechanism of light localization in waveguides with a polygonal shape of the core – cladding boundary to the ARROW mechanism. Let us consecutively consider hollow core waveguide microstructures with circular and polygonal forms of the cross – section of the core – cladding boundary wall, where the number of the polygonal sides changes in the range of $N = 3 – 8$. Waveguide properties of the polygonal microstructures were first studied in [16, 17]; in particular, the dispersion properties and loss spectra of the microstructures were investigated. It was shown that the polygonal shape of the core – cladding boundary wall led to the occurrence of asymmetrical Fano resonances in the confinement loss spectra of the air core modes.

In contrast to [16, 17] we use the above consideration to understand the basic factors of the air core mode formation in the walls of the waveguide microstructures. As it will be shown, these factors largely depend on the value of $\delta\varphi = 2\pi/N$.

It is assumed that all microstructures are made of silica glass and have the same wall thickness. The air cavities diameters were selected in such a way that the effective mode areas are approximately equal to each other in all cases. The thickness of the microstructures walls was $t = 2.76$ μm and the effective mode area in all cases $A_{eff} \approx 3900$ μm$^2$. In Fig. 1 the confinementloss dependencies on the wavelength for the fundamentals ACMs are shown in one transmission band with the center at a wavelength of 1.6 μm. The calculations were carried out by a finite element method (commercial packet Femlab 3.1) already used by us in [11]. Fig. 1 shows that the triangular waveguide microstructure has a minimal leaky loss. The pentagonal microstructure has a higher confinement loss comparable to that of the square microstructure. The octagonal microstructure has the highest confinement loss.

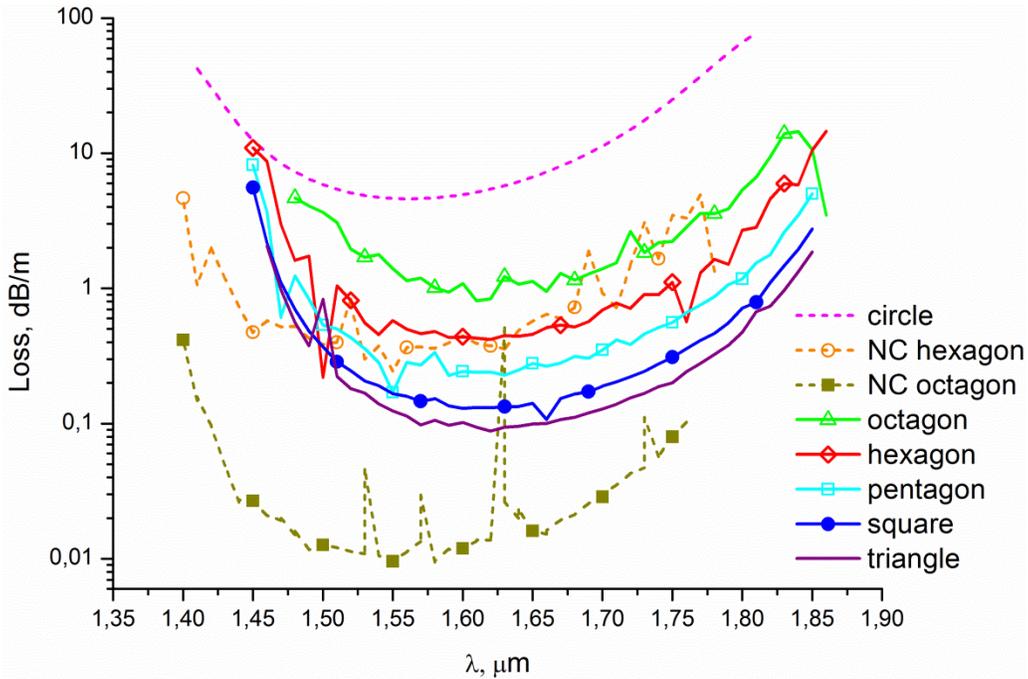

Fig. 1. Confinement loss for the waveguide microstructures with different shapes of the core – cladding boundary walls. NC denotes 'negative curvature'.

The fundamental mode of thecircular dielectric tube has a maximal value of confinement loss(Fig. 1), which is consistent with the conclusions drawn in [18]. Moreover, it can be noted that all curves in Fig. 1 have a resonant behavior inside the transmission band with the exception of those corresponding to the circular tube. Let us consequently consider the behavior of the transverse electric field components in the core - cladding boundary walls taking into account the confinement loss dependencies shown in Fig. 1.

Localization of the ACMs in the dielectric circular tube is attributed to the ARROW mechanism [5]. The distribution of the absolute value of the electric field transverse componentcommon to the ARROW mechanism is periodic in the radial direction and homogeneous in the azimuthal direction (Fig. 2). All transverse electric field distributions will be further calculatedat a wavelength of $\lambda = 1.6$ μm. The resonant condition for electric fields in the tube wall is described as in the case of the plane - parallel Fabry – Perot system by the formula $k_t d = \pi m$, where $k_{trans} = \sqrt{k_0^2 n_{silica}^2 - \beta^2}$ is the transverse component of the wavevector,$d$ is the thickness of the tube wall and $m$ is an integer. The axial dependence of the ACM is characterized by the propagation constant $\beta$.

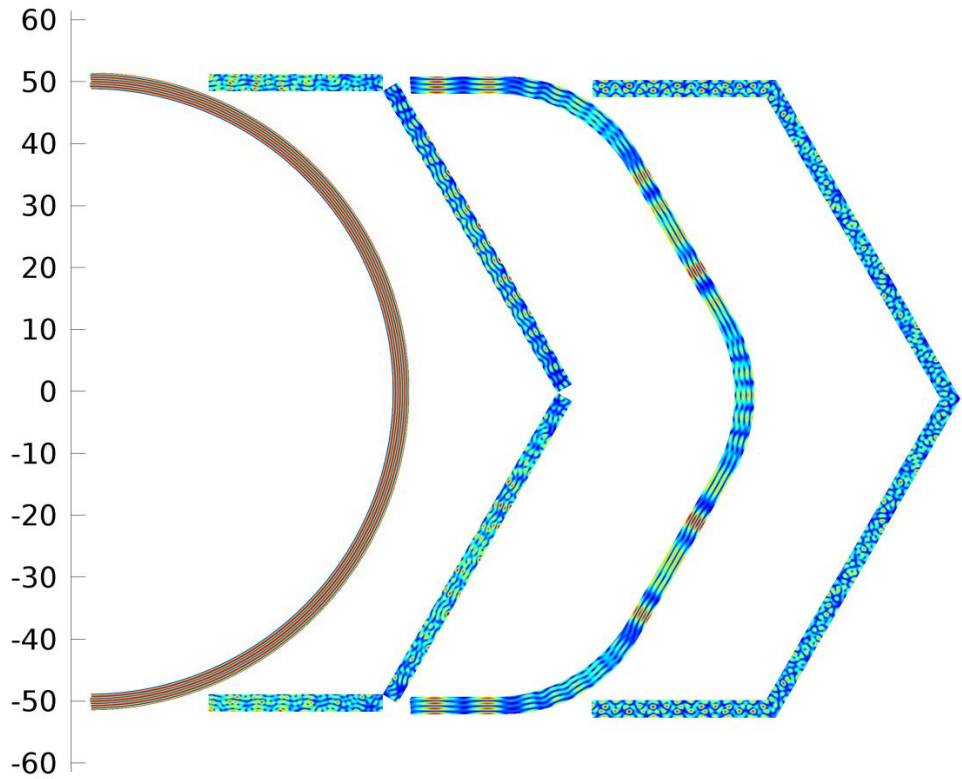

Fig. 2.The absolute value of the ACM transverse electric field components in the dielectric tube wall and in the hexagonalwaveguide microstructures wallswith different shapes of theangular domains (unbounded rectangles, smoothed angular domains and acute angular domains).

The tube wall is a good reflector in the spectral ranges where this equation is not satisfied and the antiresonant regime of radiation propagation occurs in the hollow core. Otherexamples of the ACM formation due to the ARROW mechanism are shown in Fig. 2 for the hexagonal waveguide microstructures with different shapes of the angular domains. The light is localized in the air core ofthe polygonal microstructure also according to the ARROW mechanism butthe spatial structure of electromagnetic states excited in the hexagonal microstructures walls is different from that in the dielectric tube wall (Fig. 2).The corresponding antiresonant regime of light localization in the air core leads to a different level of the waveguide loss

(Fig. 1).The transverse component of the wavevector $\vec{k}_{transv}$ can be represented as a sum of tangential and normal components in respect to each polygon side $\vec{k}_{trans} = \vec{k}_t + \vec{k}_n$. If the rectangular has the side lengths of $a$ and $b$ ($a>b$) the resonant conditions are $|\vec{k}_t|a = \pi q$ and $|\vec{k}_n|b = \pi m$, where $q$ and $n$ are integers. The two resonant conditions in the polygonal microstructure wall imply that the short wavelength band edge for the ACMs is shifted with respect to that for the ACMs in the dielectric tube (Fig. 1). For hexagonal microstructure consisting of unbounded rectangles (Fig. 2) the resonant – antiresonant conditions for the components of $\vec{k}_{transv}$ described above are fulfilled exactly.

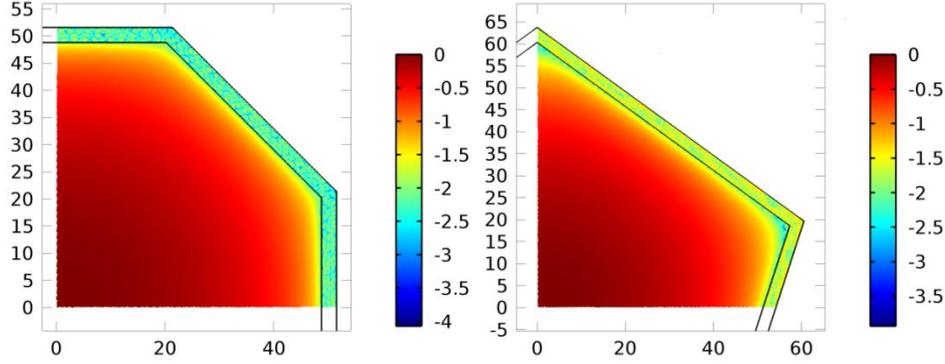

Fig. 3.The absolute value of the ACM electric field transverse component in the core – cladding boundary wall of octagonal (left) and pentagonal (right) waveguide microstructures (logarithmic scale, arbitrary units).

The situation is changed for the polygonal waveguide microstructures with a solid polygonal boundary wall shown in Fig. 2 and Fig. 3. In contrast to the ACM field distribution in the dielectric tube or in the waveguide microstructure consisting of the unbounded rectangles (Fig. 2), the light localization in the polygonal microstructure with a solid wall is more complex because the resonant – antiresonant condition for the tangential component $\vec{k}_t$ of the transverse component of the wavevector $\vec{k}_{transv}$ is disturbed due to the change in length of the outer boundary of the polygonal microstructure wall. Assuming that the propagation constants of the fundamental ACMs are approximately the same for all considered polygonal microstructures the absolute value of normal component of the transverse wavevector is $|\vec{k}_n| = \sqrt{k_0^2 n_{silica}^2 - \beta^2 - k_t^2}$ in both cases. It is clear that under disturbance of $\vec{k}_t$ in the solid microstructure wall the resonant – antiresonant condition for $\vec{k}_n$ is also disturbed. It leads to destructive interference between the transverse electric field components in the polygon wall and to a deviation from the standard ARROW mechanism of light localization. The transverse electric field components have to a large extent a disordered distribution in the solid polygonal wall, especially, in its angular domains. Depending on the number of the polygonal sides $N$ one can obtain a more or less chaotic electric field distribution in the polygonal walls at a given wavelength

(Fig. 3). Similar chaotic behavior of the electromagnetic fieldscan be observed in themicrocavities with a complicated boundary shape [19, 20]. Nevertheless, it will be shown that the difference between thelevels of waveguide loss for the hexagonal microstructures with and without angular domains shown in Fig. 2 is negligible.

Indeed, the interference of the transverse electric field components in the core – cladding boundary wall of the polygonal microstructures is largely determined by the shapes and sizes of their angular domains. In Fig. 2 the distribution of the transverse electric field components in the wall of a hexagonal microstructure with smoothed angular domainsis shown. It is seen that the regime of the destructive interference in the polygonal wall is destroyed. A steady non chaotic interference pattern in the azimuthal direction is againobservedin the microstructure wall. It is largely determined not by the individualparameters of thepolygon side but by theshape of themicrostrucutre wall as a whole, in particular, by the number of $N$ or, otherwise stated, by its discrete rotational symmetry.

The mechanism of light localization of the fundamental ACMs is quite different for microstructures with $N < 5$ ( $\delta\varphi \leq \pi/2$ ). In this case, the qualitative changes occur in the interference pattern of the transverse electric field component in the polygonal wall and the air core mode formation is not based on the same resonant - antiresonant conditions as for the above considered microstructures with $N > 4$. For the square and the triangular waveguide microstructures the distributions of the transverse electric field componentsare quite different in the central regions and in the angular domains of the core – cladding boundary walls (Fig. 4).The transverse electric field components in the angular domains have a damped character and the central regions of the microstructures sides have a field distribution that is analogous to that occurring in the dielectric tube wall (Fig. 2).Otherwise stated, the standard ARROW mechanism occurs only on limited segments of the core – cladding boundary andtends to be local.This type of light localization leads to a decrease in the waveguide loss level in comparison with the waveguide microstructures with $N > 4$ (Fig. 1). In particular,the low loss guiding at wavelength $\lambda = 2.94$ μm in silica glass triangular microstructure was demonstrated in [21].

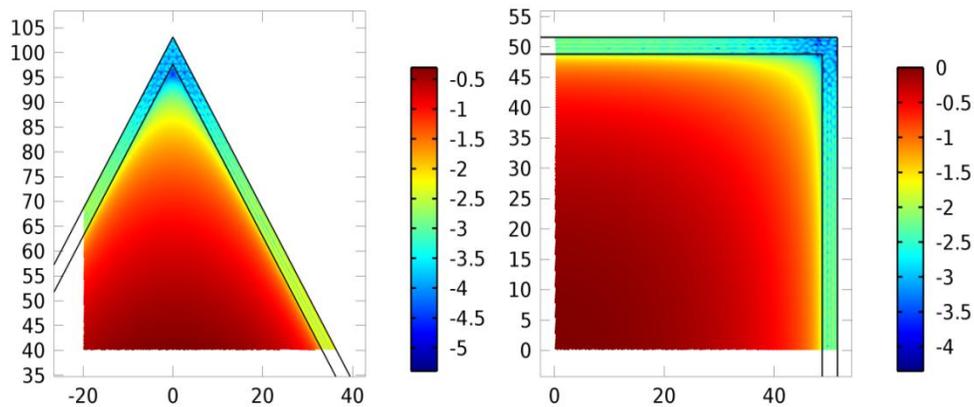

Fig. 4.The absolute value of the ACM transverse electric field component in the core – cladding boundary wall of triangular (left) and square (right) waveguide microstructures (logarithmic scale, arbitrary units).

In this way,the interference of the ACMs electromagnetic fields in the walls of the polygonal microstructures leads to a complication of boundary conditions at the core – cladding boundary and, consequently, to the spatial redistribution of the ACMs energy flowing out of the air core (Fig. 3, 4). These factors lead to a reduction in the level of waveguide loss compared to that in the circular dielectric tube (Fig. 1).Moreover,the waveguideloss for the hexagonal microstructures with acute angular domains and with the wall consisting of the unbounded rectangles (Fig. 2) isnearly identical (Fig. 5(top)).The same result can be obtained for the square microstructures (Fig. 5(bottom)).It may be concluded that the chaotic behavior of the ACMs fields in the solid core – cladding boundary wall occurring due to the presence ofangular domains doesn't significantlyaffect on the ACMs energy leakage.Astheantiresonantfield distribution for both hexagonal microstructures is different (Fig .2) and the loss level is nearly identical (Fig. 5)it is possible to assume that the factor of discrete rotational symmetry of the wall may play animportant role in the process of light localization in both cases.The transverse electric field distributions of the fundamental ACMs in the square microstructures walls are also nearly um identical and the mechanism of light localization in the air core must be the same in both cases (Fig. 5(bottom)).

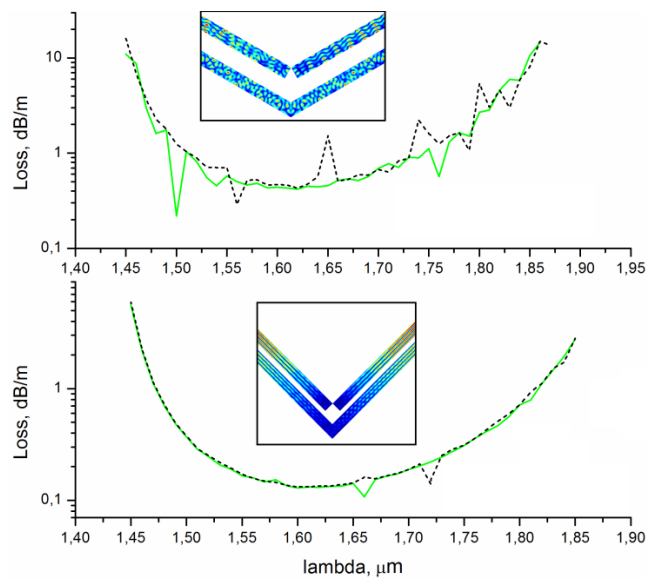

Fig. 5. The waveguide loss for hexagonal (top) and square microstructures (bottom) with (solid green) and without angular domains (black dashed).

Next, it will be shown that there exists a different way of obtaining the local ARROW mechanism in the polygonal microstructures, namely, by introducing the so called 'negative curvature' core – cladding boundary.

**III. Light localization in waveguide microstructures with a negative curvature core – cladding boundary**

There is a strong resemblance between the mechanisms of ACMs formation in waveguide microstructures with a negative curvature core – cladding boundary and those in the polygonal waveguide microstructures. Let us consider two hollow core waveguide microstructures with a negative curvature core – cladding boundary and with a different number of *N* = 6 and 8 (Fig. 6, 7).

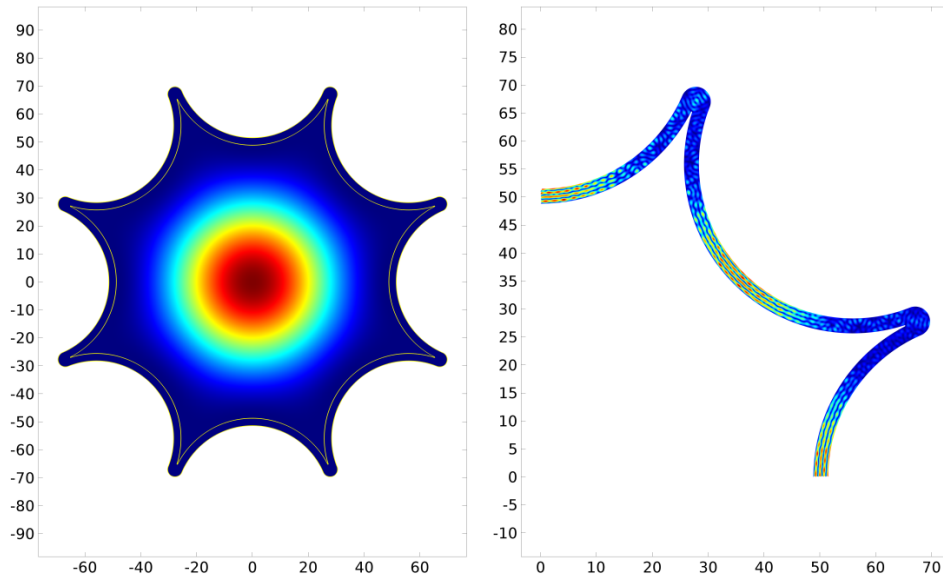

Fig. 6. The fundamental ACM of the waveguide microstrucutre with a negative curvature core – cladding boundary (*N* = 8) (left) and a distribution of the absolute value of transverse electric field component in the microstructure wall (right).

The waveguide loss of the microstructures is shown in Fig. 1. As one can see from Fig. 6(right) the fundamentalACM interacts mostly with the core – cladding boundary in the regions closest to the microstructure center. As in the case of the triangular microstructure (Fig. 4) there isa sharp fall in the transverse electric field intensity and thedisordered structure of the transverse electric field in the angular domains (Fig. 6(right)). Moreover, there is alsoa radial field distribution in the boundary regions closest to the microstructure centerwhich is formed according to the local ARROW mechanism of light localization. The waveguide loss of the fundamental ACM is several orders of magnitude lower than the loss inother polygonal microstructures (Fig. 1).

The distribution of the transverse electric field component in the microstructure wall with *N* = 6 (Fig.7) has an azimuthal periodic structure analogous to that of the hexagonal microstructure without angular domains shown in Fig. 2. For this reason, the waveguide loss in this microstructure is much higher than in the negative curvature microstructure with $N = 8$ and is comparable to the waveguide loss in the hexagonal microstructures (Fig. 1).

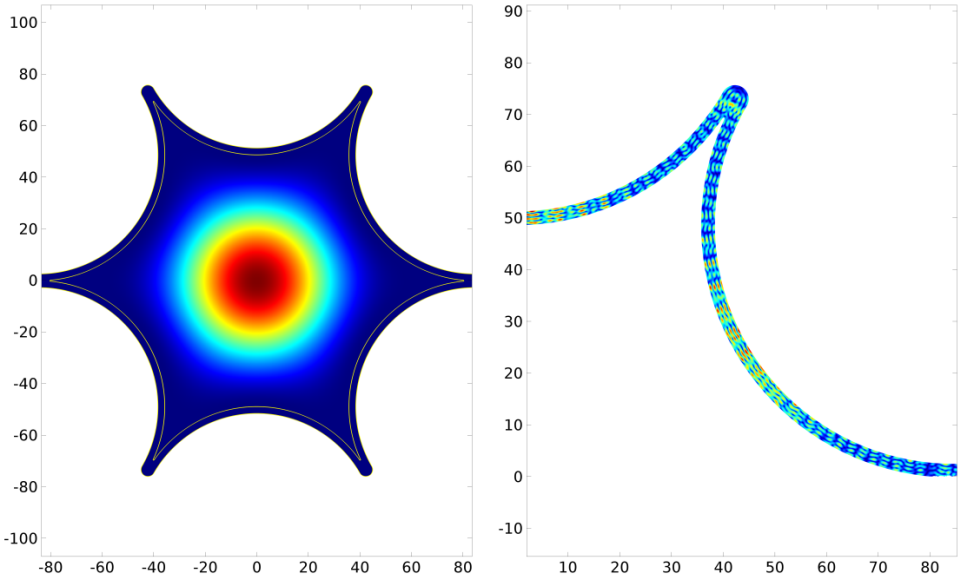

Fig. 7.The fundamental ACM of the waveguide microstructure with a negative curvature core – cladding boundary ($N = 6$) (left) and a distribution of the absolute value of the transverse electric field component in the microstructure wall (right).

Thus,the light localization in the waveguide microstructures with anegative curvature core – cladding boundary at a given wavelength can occur due to the two waveguide mechanisms that were described in Section 2. One of these mechanisms is similar to thatoccurring in the polygonal waveguide microstructures with $N > 4$. The other one is similar to the light localization mechanism in the polygonal waveguide microstructures with $N < 4$ when different parts of the polygon wall have differentACM fields distributions. In this case, the local ARROW mechanism can be applied to explaining the low lossguiding.The balance between these mechanisms can be managed by changing the negative curvature of thepolygon sides.

Let us consider in more detail the transverse electric fieldtransformation in the angular domains of the waveguide microstructure wall (Fig. 6) based on localboundary conditions that used in the theory of optical microcavities with acomplicatedboundary shape [19, 20]. Let the negative curvature core – cladding boundary be described by an equation $R = r(\varphi)$ in the polar coordinate system. Then, a vector tangent to the inner surface of the core – cladding boundary is determined by a vector product $\vec{t} = \vec{z} \times \vec{n}$, where unit vector $\vec{z}$ is directed in the axial direction and $\vec{n}(r(\varphi))$ is a unit local normal vector to the surface. Note that any vector in the core – cladding boundary region can be expressed via the vector tripod $(\vec{z}, \vec{t}, \vec{n})$ including the local transverse component of the wavevector $\left|\vec{k}_{transv}\right| = \sqrt{n_{silica}^2 k_0^2 - \beta^2}$, where $\beta$ is a propagation constant of the ACM. It is clear that the local resonant condition $|\vec{k}_n(r(\varphi))|d = \pi m$ is determined by the local value of the normal component of $\vec{k}_{transv}$, where $d$ is thickness of the microstructure wall and $m$ is an integer. In the same way it is possible to obtain the local antiresonant condition. It is clear that if the antiresonant condition at a given wavelength is fulfilled at the part of the core – cladding boundary closest to the microstructure center then the condition will change as one moves along the perimeter of the core – cladding boundary.

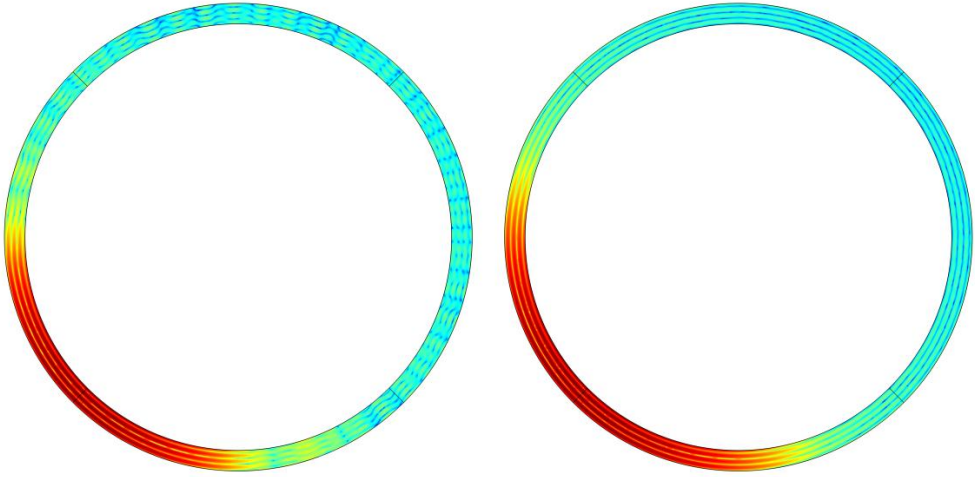

Fig.8. The distribution of the absolute values of transverse electric field components of the ACMs in the cladding capillary wall in typical HCMFs with a negative curvature core – cladding boundary: (left) the cladding capillaries almost touch each other; (right) the cladding capillaries don't touch and the distance between them is 7.7 µm. The part of the field distribution closest to the air core is red.

If the negative curvature change is relatively big at a given geometry parameters of the microstructure wall then the local tangential component $\vec{k}_t(r(\varphi))$ of the transverse wavevector grows rapidly and the antiresonant regime is only observed in small segments of the core – cladding boundary wall closest to the air core center. In this case, the local ARROW mechanism occurs (Fig. 6(right)) which is similar to that shown in Fig. 4. If it does not, then the transverse electric field distribution analogous to that in the hexagonal microstructures with a smoothed angular domain (Fig. 2) is observed. One can assume that this distribution occurs due to the absence of acute angular domains in the microstructure wall (Fig. 7) and thus a steady interference pattern is observed in the microstructure wall. The distribution of the ACM fields in the microstructure wall is also largely determined by the shape of the core – cladding boundary as a whole.

The difference in the waveguide mechanisms leads to a difference in the level of waveguide loss for the negative curvature microstructures under consideration (Fig. 1). To apply the local description of the components of $\vec{k}_{transv}$ to the entire core – cladding boundary it is necessary to consider it as continuous or quasi continuous. The vector tripod $(\vec{z},\vec{t},\vec{n})$ must characterize a behavior of the components of $\vec{k}_{transv}$ along the entire perimeter of the core – cladding boundary.

At the end of this section, let us consider the distribution of the transverse electric field component in the core – cladding boundary wall of the real NC HCMF. The NC HCMF cladding consists of eight capillaries made of silica glass. The thickness of the capillary wall is $d = 2.7$ μm, the diameter of the air core is $D_{core} = 98$ μm and the ratio of the inner diameter to the outer diameter is $d_{in}/d_{out} = 0.9$. The transverse electric field component distribution was calculated for two configurations of the NC HCMF. The first one has a very small clearance between the core – cladding boundary capillaries and the second one has a distance of 7.7 μm between the neighboring capillaries. The small clearance between the neighboring capillaries allows the ACM fields localized in the capillary wall to interact sufficiently. In this case, we can also introduce a local vector tripod $(\vec{z},\vec{t},\vec{n})$ and consider the inner surface of the core – cladding boundary as quasi continuous.

The distributions of the absolute value of the transverse electric field components of the fundamental ACM in the capillary wall are shown in Fig. 8. If the clearance between the neighboring capillaries is very small then the transverse electric field distribution is different in different regions of the capillary wall (Fig. 8(left)). The local ARROW distribution occurs in the part of the capillary wall closest to the air core center and the transverse electric field distribution similar to that in the angular domains of the microstructures with $N < 5$ (Fig. 4) is observed in the other part of the capillary wall. If the cladding capillaries are moved apart at some distance from each other then the electric field distribution in the capillary wall evolves into a standard ARROW distribution (Fig. 8(right)). As it was shown in [15] there is an optimal distance between the cladding capillaries at which the minimal waveguide loss is observed. This phenomenon can be related to the total suppression of the waveguide mechanism occurring in the polygonal microstructures walls (Fig. 3 and Fig.4) when the local vector tripod $(\vec{z},\vec{t},\vec{n})$ cannot be applied to the description of the core – cladding boundary as a whole. In this case, each cladding capillary becomes an individual scatterer with its own boundary conditions and with its own vector tripod $(\vec{z},\vec{\varphi},\vec{r})$ describing a behavior of $\vec{k}_{transv}$ in the local polar coordinate system with the origin in the center of the individual capillary. The resonant - antiresonant properties of such waveguide microstructures are well described by the standard ARROW model [5, 6].

**IV. CONCLUSIONS**

In conclusion, we have investigated the structure of the ACM fields in the core - cladding wall of different polygonal waveguide microstructures. A comparative analysis of waveguide loss of the fundamental ACMs showed that the waveguide loss in the polygonal hollow core microstructures is always lower than in the dielectric tube (the standard ARROW model) at the same wall thickness and effective mode area. It was shown that depending on the value of *N* it is possible to observe different resonant – antiresonant waveguide regimes in the polygonal walls at a given wavelength. For polygon microstructures with a number of sides $N> 4$ the standard ARROW model is disturbed and the ACM fields have to a large extent a disordered character in the microstructure wall. The polygonal microstructures with $N<5$ have a minimal loss level due to occurrence of the local ARROW mechanism of light localization. In this case, the central parts of the polygonal microstructure sides reflect the light in accordance with the standard ARROW model while the ACM field intensity in the angular domains is much weaker and the field distribution is different from that in the central parts of the sides. A similar phenomenon can be observed in NC HCFs but in this case the conditions for occurrence of the local ARROW mechanismare determined by the level of 'negative curvature' of the core – cladding boundary elements.

Note also that antiresonant reflection of light localized in the microstructured fibers air core can be achieved by two methods. The first method is well known and consists in creating a microstructured cladding which allows to localize light in the air core by using antiresonant reflection from individual elements of a microstructured cladding in the spectral range corresponding to the photonic band gap. The latter is based on creating a core – cladding boundary with a complicated shape which allows to exciteantiresonant electromagnetic field states with a complex structure in the core - cladding boundary wall. Compared to the first method, the excitation of these states can lead to a more efficient light localization in the air core.

**ACKNOWLEDGEMENT**

The reported study was supported by RFBR, research project N. 15 – 02 – 99688A.

**REFERENCES**

[1]P. St. J. Russell, "Photonic crystal fibers," J. LightWaveTechn. 24, 4729 (2006).
[2]R. F. Gregan, B. J. Mangan, J. C. Knight, T. A. Birks, P. S. J. Russell, P. J. Roberts, and D. C. Allan, "Single – Mode Photonic Band Gap Guidance of Light in Air," Science 285(5433), 1537 (1999).
[3]P. J. Roberts, F. Couny, H. Sabert, B. J. Mangan, D. P. Williams, L. Farr, M. W. Mason, A. Tomlinson, T. A. Birks, J. C. Knight, and P. St. J. Russell, "Ultimate low loss of hollow – core photonic crystal fibers," Opt. Express 13(1), 236 (2005).
[4]F. Couny, F. Benabid, P. J. Roberts, P. S. Light and M. G. Raymer, "Generation and photonic guidance of multi – octave optical frequency combs," Science 318, 1118 (2007).
[5]N. M. Litchinitser, A. K. Abeeluck, C. Headley, and B. J. Eggleton, "Antiresonant reflecting photonic crystal optical waveguides," Opt. Lett. 27, 1592 (2002).
[6]N. M. Litchinitser, S. C. Dunn, B. Usner, B. J. Eggleton, T. P. White, R. C. McPhedran, and C. Martijn de Sterke, "Resonances in microstructured optical waveguides," Opt. Express 11, 1243 (2003).
[7]A. Argyros and J. Pla, "Hollow core polymer fibres with a Kagome lattice: potential for transmission in the infrared," Opt. Express, 15, 7713 – 7719 (2007).
[8]A. Argyros, S. G. Leon – Saval, J. Pla and A. Docherty, "Antiresonant reflection and inhibited coupling in hollow - core square lattice fiber," Opt. Express, 16, 5642 – 5648 (2008).
[9]Y. Y. Wang, N. V. Wheeler, F. Couny, P. J. Roberts, and F. Benabid, "Low loss broadband transmission in hypocycloid - core Kagome hollow – core photonic crystal fiber," Opt. Lett., 36, 669 (2011).

[10]B. Debord, M. Alharbi, T. Bradley, C. Fourcade – Dutin, Y. Y. Wang, L. Vincetti, F. Gerome, and F. Benabid, "Hypocicloid – shaped hollow core photonic crystal fiber: I. Arc curvature effect on confinement loss," Opt. Express 21, 28597 (2013).

[11]A. D. Pryamikov, A. S. Biriukov, A. F. Kosolapov, V. G. Plotnichenko, S. L. Semjonov, and E. M. Dianov, "Demonstration of a waveguide regime for a silica hollow – core microstructured optical fiber with a negative curvature of the core boundary in the spectral range > 3.5 μm," Opt. Express 19, 1441 (2011).

[12]F. Yu, W. Wadsworth, and J. C. Knight, "Low loss silica hollow core fiber for 3 – 4 μm spectral range," Opt. Express 20, 11153 (2012).

[13]A. N. Kolyadin, A. F. Kosolapov, A. D. Pryamikov, A. S. Biriukov, V. G. Plotnichenko, and E. M. Dianov, "Light transmission in negative curvature hollow core fiber in extremely high material loss region," Opt. Express, 21, 9514 – 9519 (2013).

[14]W. Belardi and J. C. Knight, "Effect of core boundary curvature on the confinement losses of hollow core antiresonant fibers," Opt. Express 21, 21912 (2013).

[15]G. K. Alagashev, A. D. Pryamikov, A. F. Kosolapov, A. N. Kolyadin, A. Yu. Lukovkin, and A. S. Biriukov, "Impact of geometrical parameters on the optical properties of negative curvature hollow – core fibers," Laser Phys. 25, 055101 (2015)

[16]L. Vincetti and V. Setti, "Fano resonances in polygonal tubes fibers," J. Lightwave Tech. 30, 31 – 37 (2012).

[17]L. Vincetti and V. Setti, "Extra loss due to Fano resonances in inhibited coupling fibers based on a lattice of tubes," Opt. Express, 20, 14350 – 14361 (2012).

[18]W. Ding and Y. Wang, "Semi analytical model for hollow core antiresonant fibers," Front. Phys., 3, article 16 (2015).

[19]H. Cao and J. Wiersig, "Dielectric microcavities: model systems for wave chaos and non – Hermitian physics," Rev. Mod. Phys. 87, 61 (2015).

[20]H. E. Tureci, H. G. L. Schwefel, Ph. Jacquod and A. Douglas Stone, "Modes of wave – chaotic dielectric resonators," Progress in Optics, 47, 75 – 137 (2005).

[21]Y. Chen, M. F. Saleh, N. Y. Joly, and F. Biancalana, "Guidung 2.94 μm using low – loss microstructuredantiresonant triangular – core fibers," Journal of Applied Physics, 119, 143104 (2016).